\documentclass[preprint,showkeys,showpacs,onecolumn,nofootinbib,notitlepage,superscriptaddress,longbibliography]{revtex4-2}

\usepackage[colorlinks=true,
            linkcolor=blue,
            urlcolor=blue,
            citecolor=blue]{hyperref}
        
\usepackage{amssymb,amsmath,epsfig}
\usepackage{eurosym}
\usepackage{amsfonts}
\usepackage{array}
\usepackage{changes}
\usepackage{amsthm}

\def\0{{\sst{(0)}}}
\def\1{{\sst{(1)}}}
\def\2{{\sst{(2)}}}
\def\3{{\sst{(3)}}}
\def\4{{\sst{(4)}}}
\def\5{{\sst{(5)}}}
\def\6{{\sst{(6)}}}
\def\7{{\sst{(7)}}}
\def\8{{\sst{(8)}}}
\def\sst#1{{\scriptscriptstyle #1}}

\usepackage{mathpazo}
\usepackage{supertabular}
\usepackage{graphics}
\usepackage{color}
\usepackage{graphicx}

\usepackage{caption}
\usepackage{subcaption}
\usepackage{orcidlink}
\begin{document}

\title{Thin accretion disk images of the black hole in symmergent gravity}

\author{{\.I}lim {\.I}rfan {\c C}imdiker}
\email{ilim.cimdiker@phd.usz.edu.pl}
\affiliation{Institute of Physics, University of Szczecin,
Wielkopolska 15, 70-451 Szczecin, Poland}

\author{Ali {\"O}vg{\"u}n}
\email{ali.ovgun@emu.edu.tr (Corresponding Author)}
\affiliation{Physics Department, Eastern Mediterranean University, Famagusta, 99628 North Cyprus, via Mersin 10, Turkey}

\author{Durmu{\c s} Demir}
\email{durmus.demir@sabanciuniv.edu}
\affiliation{Faculty of Engineering and Natural Sciences, Sabanc{\i} University, 34956 Tuzla, {\.I}stanbul, Turkey}

\begin{abstract}

In this paper, we study circular orbits, effective potential, and thin-accretion disk of a black hole in symmergent gravity within the Novikov-Thorne model in a way including the energy flux and temperature distribution. We determine bounds on symmergent gravity parameters and conclude that the accretion disk could be used as an astrophysical tool to probe symmergent gravity. 
\end{abstract}
\date{\today}


\maketitle

\section{Introduction}
Since 1915 General Relativity(GR) has stood firm against many tests and stands as one of the most successful theories of modern physics. First, the theory held out against Eddington's test around the Sun's bending of light at low energy regimes. Later on, observation of the gravitational waves at the weak field limit proved GR successful. Recently, observations of super-massive black holes M87$^*$ and the Sgtr $A^*$ have been the primary focus for many gravity theorists. Black holes are objects containing a real singularity at the center encased by a coordinate singularity, which is the horizon \cite{EventHorizonTelescope:2019dse, EventHorizonTelescope:2019ggy, Akiyama:2022tyh, Akiyama:2022yot, Akiyama:2022qhc}. They are anomalous objects. Tests around them would probe new sectors of modern physics.

Besides the GR, a non-renormalizable theory at its core, we only have a spontaneously broken gauge theory, the Standard Model(SM), to explain our universe. The main problem of modern physics is that we can not merge these two theories and obtain a working unified model. Directly quantizing gravity yields divergences that cannot be cured, and the prominent string theory is troubled with high string tensions and the absence of super-symmetry \cite{dyson, thooft, incompatible, weinberg, eft}.

Some frameworks define gravity as not fundamental but an effective theory that results from the underlying phenomena. Sakharov's Induced gravity, for instance, derives an approximate Newton constant from the loop coefficient of a scalar field action that is minimally coupled to the gravity \cite{sakharov}. These kinds of theories are called induced or emergent gravity theories. In such frameworks, gravity is not considered fundamental but rather an emergent phenomenon that bounds the underlying theory \cite{visser,verlinde}.

Also, a novel framework called Symmergent Gravity(SG) extends the SM by inducing the required curvature sector from the loop coefficient of the effective action of an underlying Quantum Field Theory(QFT) by requiring gravity to emerge in a fashion that restores the broken gauge symmetries \cite{demir1,demir2,demir3, irfan, Cimdiker:2021cpz}. It contains standard linear scalar curvature $R$ beside a quadratic curvature term $R^2$ coupled to loop coefficients that the underlying QFT can adjust \cite{birrel}. The loop coefficient is nullified under a super-symmetric structure of the new physics sector, whereby leading to the Einstein-Hilbert action. Also, Since the theory has an $R+R^2$ structure similar to $f(R)$ gravity theories \cite{f(R)review1,Clifton:2011jh,DeFelice:2010aj,Olmo:2011uz,Cembranos:2011sr,Nashed:2019tuk,Elizalde:2020icc,Nashed:2021sey,Zheng:2018fyn,Moon:2011sz,Myung:2011ih}, the theory can lead to Starobinsky inflation under a boson heavy system. 

The recent result about the M87$^*$ and Sgtr $A^*$ from the Event Horizon Telescope(EHT) gives a clear insight into black holes, and the results can be a test-bed for many modified theories of gravity. The observed image of the M87$^*$ and Sgtr $A^*$ reflects a so-called shadow of the black hole, representing the bent light rays from the photon sphere and the heated accretion disc of the black hole to a faraway observer. This shadow cast and the simulated images of the thin accretion disc should constraint the modification of GR and has been studied in many articles \cite{Perlick:2015vta,Luminet:1979nyg,Falcke:1999pj,Ovgun:2018tua,Ovgun:2020gjz,Okyay:2021nnh,Pantig:2021zqe,Pantig:2022whj,Pantig:2022toh,Pantig:2022qak,Rayimbaev:2022hca,Pantig:2022ely,Uniyal:2022vdu,Gralla:2019xty,Takahashi:2004xh,Allahyari:2019jqz,Dokuchaev:2020wqk,Zeng:2020dco,Paul:2019trt,Johannsen:2015mdd,Johannsen:2016vqy,Johannsen:2015hib,Broderick:2016ewk,Rahaman:2021web,Amir:2018szm,Shaikh:2021cvl,Shaikh:2018lcc,Shaikh:2019fpu,Khodabakhshi:2020hny,Wei:2019pjf,Hennigar:2018hza,Abdolrahimi:2015kma,Khodadi:2021gbc,Khodadi:2020jij,Vagnozzi:2022moj,Chen:2022nbb,Roy:2021uye,Vagnozzi:2020quf,Vagnozzi:2019apd,Heydari-Fard:2021ljh,Afrin:2021imp,Kumar:2020hgm,Kumar:2018ple,Kumar:2020owy,Atamurotov:2013sca,Tang:2022hsu,Kuang:2022ojj,Kuang:2022xjp,Meng:2022kjs,Bambi:2015kza}. 

Recently, black hole solutions of the SG framework have been found, and a detailed investigation of its shadow, angular radius, and thermodynamics has been done \cite{Cimdiker:2021cpz}. In this paper, we would like to generalize our findings by finding circular orbits for massive test particles and constraint SG. Since SG is tied to the underlying QFT, black hole properties would help us probe the new physics sector (predicted by the SG framework). 

The structure of the paper is as follows. In section II, we give a brief description of the black hole structure in SG theory. Section III finds proper circular orbits and effective potentials in SG for Schwarzschild and Kerr solutions. Section IV considers an SG black hole with a thin accretion disk. Finally, in section V we conclude.

\section{Black hole in Symmergent Gravity}

Symmergent gravity is a novel idea and shorthand name for ``gauge symmetry- restoring emergent gravity". (The setup, the problem, and all the mechanism details can be found in the recent work \cite{demir1}.)

The most elusive problem with the established framework of physics is the unification of gravity and the SM. The consensus is that the SM is incomplete for various reasons, such as dark energy and dark matter. The idea in SG is that effective QFTs can be completed in the UV in analogy to the Higgs mechanism, with the main difference that instead of the Higgs field once adopts curvature in view of the fact that the UV cutoff breaks Poincare symmetry (like the curvature). By harnessing the symmetry properties between the UV scale in a flat manifold and the scalar curvature $R$ or $\mathcal{R}$ (metric or affine) in a curved manifold, SG constructs an effective action by mapping the UV scale to an affine curvature $\mathcal{R}$. The solution to the framework above yields the following curvature sector.

\begin{eqnarray}
\label{curvature-sector}
\int d^4x \sqrt{-g}\left\{
\frac{R(g)}{16\pi G_N} - \frac{c_o}{16} R(g)^2\right\}
\end{eqnarray}
where $R(g)$ is the standard metric curvature, $c_o$ and $G_N$ are flat spacetime loop-induced coefficients: 

\begin{eqnarray}
\label{params}
\frac{1}{G_N} = \frac{{\rm str}\left[m^2\right]}{8 \pi}\,,\;c_o =-\frac{(n_b-n_f)}{64 \pi^2} 
\end{eqnarray}
in which $n_b (n_f)$ is the total number of bosonic  (fermionic) degrees of freedom in the underlying QFT. From eq.(\ref{curvature-sector}), one can see that the result of SG contains both linear and quadratic curvatures accompanied by loop coefficients. Such models can be represented by $f(R)$ gravity, where instead of a standard Einstein-Hilbert action with a linear curvature $R$, the theory contains a function $f(R)$ of the curvature in the form

\begin{eqnarray}
S=\frac{1}{16 \pi G_N} \int d^4 x \sqrt{-g}  f(R)
\label{fr-action}
\end{eqnarray}

so that the SG action in (\ref{curvature-sector}) can be recast as an $f(R)$ gravity action as in eq.(\ref{fr-action}) by picking the following functional \cite{Cimdiker:2021cpz}:

\begin{eqnarray}
\label{fR}
f(R)= R - \pi G_N c_o R^2 
\end{eqnarray}

The quadratic term $R^2$ is accompanied by the loop coefficient $c_o$, which is directly tied to the number difference between bosonic and fermionic degrees of freedom forming the QFT spectrum. Since the quadratic term is the leading correction to the EH action, many anomalies can be tackled by fine-tuning its loop coefficient. For instance, the numbers $n_b$ and $n_f$ can take specific values to explain the early Starobinsky inflation as it is an $R+R^2$ gravity theory (a special case of $f(R)$ theory). Furthermore; the equation of motion for eq.(\ref{fR}) reads as

\begin{eqnarray} 
\label{f1}
\mathit{R}_{\mu \nu} \mathit{F(R)}-\frac{1}{2}g_{\mu \nu}\mathit{f(R)}+[g_{\mu \nu}\Box -\nabla_\mu \nabla_\nu]\mathit{F(R)}= 0
\end{eqnarray}
in which
\begin{eqnarray}
\label{FR}
F(R) = \frac{d f(R)}{d R} = 1 - 2 \pi G_N c_o R
\end{eqnarray}
is the derivative of $f(R)$ in eq.(\ref{fR}) with respect to the its argument. Making use of its trace, eq.(\ref{f1}) can be recast as a traceless motion equation
\begin{eqnarray} 
\label{f3ss}
0 &=& \mathit{R}_{\mu \nu} \mathit{F(R)}-\frac{g_{\mu \nu}}{4}\mathit{ R}\mathit{F(R)}+\frac{g_{\mu \nu}}{4}\Box\mathit{F(R)} -\nabla_\mu \nabla_\nu\mathit{F(R)} \nonumber \\
\newline
\end{eqnarray}
solutions of which will have an explicit dependence on the loop factor $c_o$. Thus, it varies with the particle content of the underlying QFT. Furthermore, for a static, spherically symmetric metric 
\begin{eqnarray} \label{met12}
ds^2=-B(r)dt^2+\frac{dr^2}{B(r)}+r^2(d\theta^2+\sin^2\theta d\phi^2)\,  
\end{eqnarray}
parametrized by the radial coordinate $r$, angular coordinates $\theta$ and $\phi$, and yet-to-be-determined lapse function $B(r)$, the scalar curvature takes the form

\begin{equation} 
\label{scalar-curve} 
R={\frac { -B'' {r}^{2}-4\,r B'
 -2\,B  +2}{{r}^{2}}}
\end{equation}
Furthermore, inserting scalar curvature (\ref{scalar-curve}) into eq.(\ref{f3ss}) yields the following equations (equations in $\theta$ and $\phi$ directions are identical up to a $\cos\theta$ factor)
\begin{flalign}
\label{E1n}
2FB'' + 2B'F'
- 2BF'' - \frac{4}{r}BF' + \frac{4}{r^2}F (1 - B) &=0, \\
\label{E2n}
2FB'' + 2B'F'
 + 6BF'' -\frac{4}{r}B F' + \frac{4}{r^2}F (1 - B) &=0, \\
\label{E3n}
2FB'' + 2B'F'+ 2BF'' - \frac{4}{r}BF' + \frac{4}{r^2}F (1 - B) &=0 
\end{flalign}
These three equations consecutively lead to $B F''  = 0$. Thus one can construct a linear solution for $F(r)$ as 
\begin{eqnarray}
\label{FRsoln0}
F[R(r)] = \xi_1 + \xi_2 r 
\end{eqnarray}
with $\xi_1 $ and $\xi_2$ undetermined constants. One can find a proper lapse function if and only if $\xi_2=0$ reduces eqs. (\ref{E1n},\ref{E2n},\ref{E3n}) to
\begin{eqnarray}
    2 F B'' +\frac{4}{r^2}F(1-B)=0
\end{eqnarray}
Thus one can find the lapse function as,
\begin{eqnarray}
\label{newmet}
B(r)=1-{\frac {{\mu}}{r}}-{\frac{ r^2(1-\xi_1)}{24 \pi G_N\,c_o}}
\end{eqnarray}
The lapse function includes a quadratic radial part directly tied to the loop coefficient $c_o$ (which also contains the integration constant $\xi_1 $ as a free parameter\footnote{In the following discussions, we will take the free parameter to be the order of null. Notice that $\xi_1\rightarrow 1$ limit yields Schwarzchild case; thus, making it null would not affect the results.}). As the loop coefficient, $c_o$, increases, the lapse function approaches the standard Schwarzschild solution with the integration constant set to $\mu=2 G_N M$. This is a Schwarzschild-AdS or Schwarzschild-dS solution depending on the loop factor (and the free parameter $\xi_1$), and we will see that the $c_o$ will play the part of a cosmological constant\footnote{This is expected due to the following. $F(R)$ theories can be cast out from the Einstein frame to the Jordan frame to obtain a scalar-tensor theory, where the functional $F(R)$ plays the role of the scalar field. Thus non-asymptotically flat solutions of such a theory would lead to a dS or AdS solution for black holes. On the other hand, Asymptotically flat solutions for $\xi_1=0$ would lead to the standard Schwarzchild solution.}. Furthermore, one can calculate the horizon radius as
\begin{eqnarray}
\label{horizon}
r_{h}= \frac{h}{(18)^{1/3}} -\frac{(18)^{1/3} c_1}{3 h} 
\end{eqnarray}
where
\begin{eqnarray}
h = ((12 c_1^3 + 81 c_2^2)^{1/2} - 9 c_2)^{1/3}
\end{eqnarray}
with $c_1=1/3\Lambda$, $c_2=\mu/3\Lambda$, $\Lambda=(1-\xi_1)/(8\pi G_N c_o)$. 

To summarise, SG is a higher-order emergent gravity theory that results in an AdS or dS-type black hole solution depending on the loop coefficient of the underlying QFT.

\section{Motion of Test Particles}

In this section, we will investigate the motion of a test particle in symmergent gravity in the Schwarzchild and Kerr geometries. In this regard, we will follow the methodology in \cite{Perez:2012bx,Harko:2009xf}.

\subsection{Motion Around Schwarzchild BH in SG}
\label{subsec-1}
\begin{figure}[!ht]
    \centering
    \includegraphics[scale=0.6]{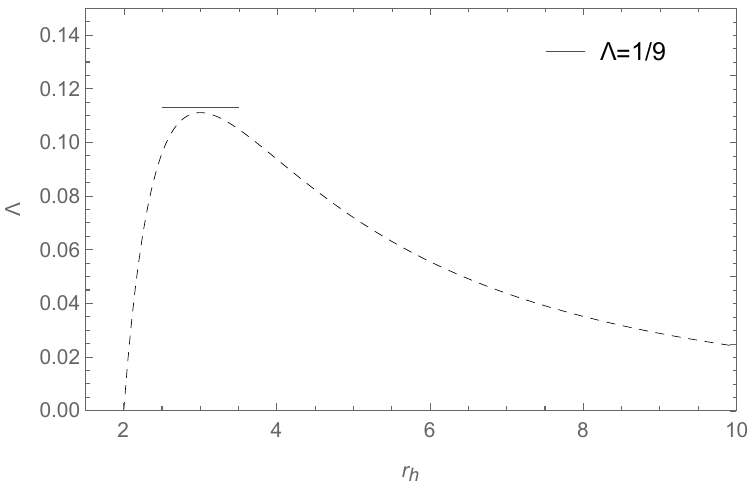}
    \caption{Limit on $\Lambda$ at the horizon. Beyond $4/9\mu^2$ ($\Lambda=1/9$ for $G_N=M=1$) naked singularities occur.}
    \label{fig:horizonsc}
\end{figure}

Depending on the loop coefficient $c_o$, SG admits Schwarzchild-dS ($\Lambda>0$) and Schwarzchild-adS ($\Lambda<0$) spacetimes, which are determined by the following static, sperically-symmetric line element
\begin{eqnarray}
    ds^2=-B(r) dt^2 + B(r)^{-1} dr^2+ r^2 (d\theta^2+\sin^2\theta d\phi^2) \label{static} \\ 
    B(r)=1- \frac{\mu}{r}-r^2\frac{\Lambda}{3} \\
    \mu=2 G_N M\;\;\;\; \textit{and}\;\;\;\;\; \Lambda(c_o) =\frac{1-\xi_1 }{8 \pi G_N c_o}.
\end{eqnarray}
Horizon can be found by setting $B(r)=0$ and solving for $r$. Solving for $\Lambda$ yields $\Lambda=3(r_h-\mu)/r_{h}^3$ where $r_h$ is the horizon of the black hole. We plotted this relation in Fig. \ref{fig:horizonsc} where we see the maximum of $\Lambda$ occur at $\Lambda_{max}=4/9\mu^2$ ($\Lambda=1/9$ for $G_N=M=1$). Beyond this limit, naked singularities occur. Thus for a positive loop coefficient $c_o$, we have an upper bound for $c_o$, and two horizons occur (inner and outer horizons). This bound corresponds to $c_o= [9(1-\xi_1 )/8 \pi ] \times(G_N M^2)$ for loop coefficient. For negative $c_o$, we have only one horizon and no limit on parameter.\footnote{The loop-factor $c_o$ is a dimensionless parameter leading to the effective cosmological constant $\Lambda$. The bound on $c_o$ depends on the mass of the black hole. This can be understood as follows, for a given system with fixed $c_o$ (that is, a fixed difference between the numbers of degrees of freedom between bosons and fermions), a naked singularity can occur if this bound is violated as $M_{bh}>\sqrt{9(1-\xi_1)/8 \pi c_o G_N}$ where $M_{bh}$ is the mass of the black hole.}

The geodesic equation governs the motion of test particles. Defining the four momenta  $p^\mu=\Dot{x}^\mu$ where the dot stands for derivative with respect to an affine parameter (proper time for massive particles) and normalizing $p^\mu p_\mu=-1$ (test particles of unit mass), the conjugate momenta are found to be $P= B(r) \Dot{t}$ and $\Phi=r^2 \Dot{\phi}$ where $P$ and $\Phi$ are linear and angular momenta of the particle. In the equatorial plane ($\theta =\pi/2$), motion of the test particle is governed by the effective potential 
\begin{eqnarray}
    V_{eff}= B(r)\left(m + \frac{\Phi^2}{r^2}\right)
\end{eqnarray}
so that the equation of motion reads as $\Dot{r}^2=P^2-V_{eff}$. The motion is seen to be allowed in regions where $P^2>V_{eff}$ and the radial motion can be described as $\Dot{r}=0$.
\begin{figure}[!ht]
    \centering
    \includegraphics[scale=0.6]{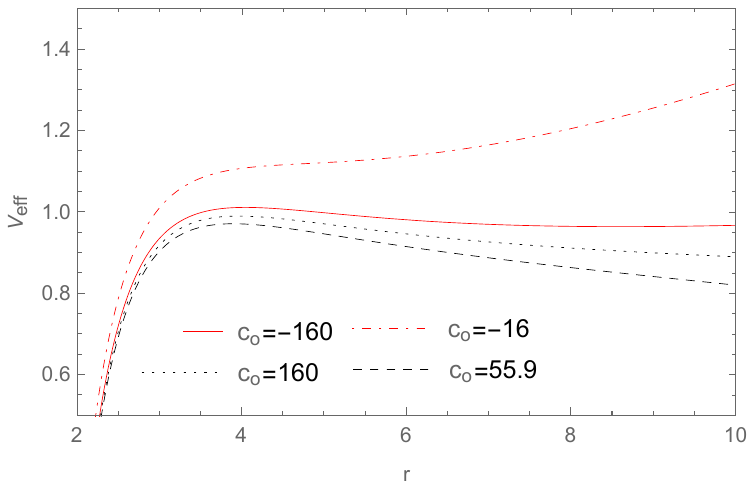}
    \caption{Effective potential $V_{eff}$ as a function of the radius $r$ for several $c_o$ values with $G_N=M=1$ and $\Phi=4$.}
    \label{fig:veffsch}
\end{figure}

In Fig. \ref{fig:veffsch}, we depict the profile of $V_{eff}$ for various positive and negative $c_o$ values. The circular orbits correspond to local extrema ($V_{eff}'(r)=0$), and concavity determines the stability of the orbit (for stable orbits $V_{eff}''(r)>0$). One can solve $V_{eff}'=0$ for $\Phi$ to get the  angular momentum on the orbit as
\begin{eqnarray}
    \Phi=\sqrt{\frac{r_c}{3} \frac{ (2r_c^3 \Lambda - 3 \mu)}{\left(\frac{3 \mu}{r_c}-2\right)}}  \label{Phi}  
\end{eqnarray}
where $r_c$ is the corresponding circular orbit. Thus circular orbits can exist within the limit of $3\mu/2<r_c<(3 \mu/2 \Lambda)^{1/3}$. At $r=3 \mu/2$, orbit radii remain below by the stable circular photon orbit, which lies at $r=3G_N M$.

For $\Lambda>0$, the radii are bound to lie above by $(3 \mu/2 \Lambda)^{1/3}$ as ``static radius" \cite{Stuchlik:1999qk} for which pressure from $\Lambda$ negates the gravitational attraction. With the specific angular momenta values we obtain from (\ref{Phi}), we solve $V''(r)=0$ for $\Lambda$ to find
\begin{eqnarray}
    \Lambda(c_o)= \frac{3 \mu (r_c-3 \mu)}{r_c^3 (8 r_c -15 \mu)} \label{Lambda}
\end{eqnarray}
in which the radius $r=r_c$ corresponds to marginally stable orbits. We plot this parametric relation in Fig. {fig:lambdavsrsch}. 
\begin{figure}[!ht]
    \centering
    \includegraphics[scale=0.6]{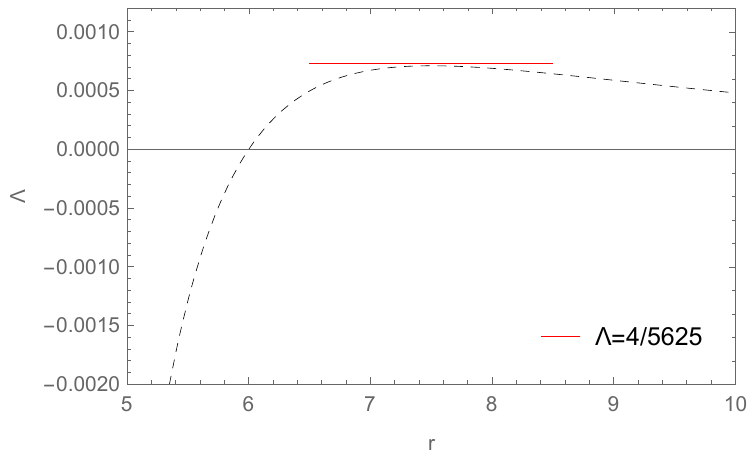}
    \caption{The parametric relation (\ref{Lambda}). Here, the red tangent line corresponds to maximum values of $\Lambda$ for which a stable circular orbit forms ($G_M=M=1$).}
    \label{fig:lambdavsrsch}
\end{figure}
At $r_c=3 \mu$, we recover the Schwarzchild orbit (without cosmological constant). The maxima of the equation (\ref{Lambda}) is at $r_c=15 \mu /4$ which puts an upper bound on $\Lambda$ as $\Lambda<16/5625 \mu^2$($\Lambda=4/5625$ for $G_N=M=1$). This means that in the SG framework, a gravitational object will have a circular orbit if and only if its mass is below critical mass $M_{cr}$ which follows
\begin{eqnarray}
    \frac{5625}{32\pi}G_N M^2_{cr}<c_o\,. \label{limit}
\end{eqnarray}

For de Sitter solutions, stable circular orbits can form for all values of $\Lambda$ and hence $c_o$. In this case, the radius has no upper limit ($r_{osco}$). Stable orbits can again be determined via the concavity condition in eequation (\ref{Lambda}). In the table (\ref{tab:risco}), we have numerically calculated innermost and outermost circular orbits for various $c_o$ values (in units of $G_N M^2)$) and $\Tilde{c}_o$ for the (hypothetical non-rotating) mass of Cyg X-1.   
\begin{table}[!ht]
    \centering
    \begin{tabular}{ p{1.5cm} p{1cm} p{1cm} p{1.5cm} p{2cm} }
        \hline
        $\Lambda(1/\mu^2)$ & $r_{isco}(\mu)$ & $r_{osco} (\mu)$ & $c_o(G_N M^2)$ & $\Tilde{c}_o(CygX1)$  \\
        \hline
        \hline
        $2.8 \times 10^{-3}$ & $3.7$ & $3.8$& \; $55.9$&  $9.6\times 10^{79}$\\
        $\;\;10^{-6}$ &  $3$ & $71.7$&    $1.6\times10^{5}$&  $2.7\times 10^{83}$\\
        $\;\;10^{-12}$ &  $3$ & $7210$&    $1.6\times10^{11}$&  $2.7 \times 10^{89}$\\
        $-10^{-3}$ &  $2.92$ & $-$&    $-160$&  $-2.7 \times 10^{80}$\\
        $-10^{-2}$ &  $2.63$ & $-$&    $-16$&  $-2.7 \times 10^{79}$\\
        \hline
        
    \end{tabular}

    \caption{Stable orbits for symmergent dS-AdS Schwarzchild black holes for various values of the loop factor $c_o$. Here, $\mu=2G_N M $ and $\Tilde{c_o}$ correspond to the X-Ray source $Cyg X-1$ with $M_{CygX1}=1.6 \times 10^{67} eV/c^2$. The upper limit on $\Tilde{c}_o$ to form circular orbits, calculated from equation (\ref{limit}),  reads to be $9.6\times 10^{79}$.}
    \label{tab:risco}
\end{table}

\subsection{Motion Around Kerr BH in SG}
\label{subsec-2}

In symmergent gravity, the stationary metric describing a rotating black hole of mass $M$ and angular momentum $a$ takes the form
\begin{eqnarray}
ds^2 = \frac{\sigma^2}{\mathcal{B}_r} dr^2+ \frac{\sigma^2}{\mathcal{B}_\theta} d\theta^2
+\frac{\mathcal{B}_\theta \sin^2 \theta}{\sigma^2}\left( \frac{a}{\chi} dt-\frac{(r^2+a^2) }{\chi}d\phi\right)^2 \nonumber\\
-\frac{\mathcal{B}_r}{\sigma^2}\left(\frac{1}{\chi}dt-\frac{a\sin^2 \theta}{\chi} d\phi  \right)^2 \;\;\;\;\;\;\;\; 
\label{metric-0}
\end{eqnarray}
in which various parameters are found to be
\begin{eqnarray}
\chi &=& 1+\frac{\Lambda}{3}a^2 \\
\sigma &=& \sqrt{r^2 + a^2 \cos \theta^2} \\
\mathcal{B}_r &=& (r^2+a^2)(1-\frac{\Lambda}{3}r^2)- \mu r \\
\mathcal{B}_\theta &=& 1+\frac{\Lambda}{3} a^2 \cos^2 \theta
\end{eqnarray}
after solving the motion equations. Here, $\chi$ depends on the loop factor $c_o$ through $\Lambda(c_o)$ given in equaton (\ref{Lambda}). In the equatorial plane (i.e. $\theta=\pi/2$ ), the metric (\ref{metric-0}) reduces to
\begin{eqnarray}
ds^2= -\frac{(\mathcal{B}_r-a^2)}{r^2 \chi^2}dt^2 +\frac{r^2}{\mathcal{B}_r}dr^2
-\frac{2 a (r^2+a^2-\mathcal{B}_r)}{r^2 \chi^2} dt d\phi\nonumber\\ + \frac{((r^2+a^2)^2-\mathcal{B}_r a^2)}{r^2 \chi^2}d\phi^2 \;\;\;\;\;\;\;\;\label{equatorialmetric}
\end{eqnarray}
from which the horizon radius $r_h$ is determined by solving $\mathcal{B}_r=0$ for $r$. Meanwhile, $\Lambda$ is related to the horizon radius through the relation
\begin{eqnarray}
    \Lambda=\frac{3 \left(a^2+r_h^2-\mu  r_h\right)}{r_h^2 \left(a^2+r_h^2\right)} \label{lambdakerr}
\end{eqnarray}
\begin{figure}[!ht]
    \centering
    \includegraphics[scale=0.6]{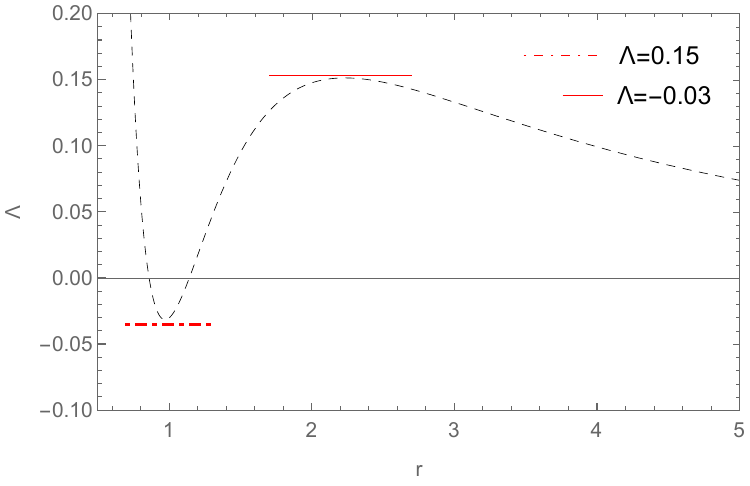}
    \caption{The parametric relation in equation (\ref{lambdakerr}). Here, red tangents stand for maximum (full line) and minimum (dashed line) values of $\Lambda$ for which a circular orbit forms for a maximally rotating black hole ($a=0.99$).}
    \label{fig:horizonkerr}
\end{figure}

We plot in Fig. \ref{fig:horizonkerr} the parametric relation in equation (\ref{lambdakerr}) for a maximally rotating black hole ($a=0.99$). Two critical points correspond to $\Lambda_1=0.15\times (1/G_N^2 M^2)$ and $\Lambda_2=-0.03 \times (1/G_N^2 M^2)$. In the range $0<\Lambda<\Lambda_1$, there arise 3 singularities: the inner, outer, and the cosmological horizons. In the case of $\Lambda_2<\Lambda<0$, one the other hand, there arise 2 singularities. Finally, for $\Lambda_1<\Lambda$ there arises a single horizon. For $\Lambda$ below $\Lambda_1$, naked singularities occur and this critical value corresponds to  $c_o=-1.32\times (G_N M^2)$. 

\begin{figure}[!ht]
    \centering
    \includegraphics[scale=0.6]{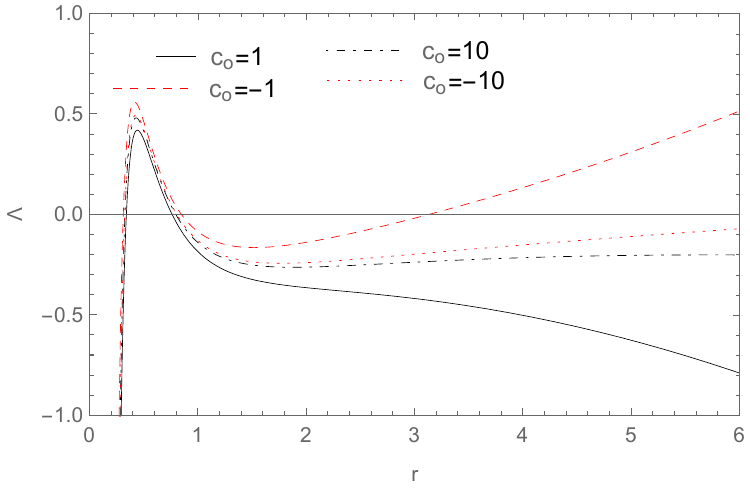}
    \caption{Variation of the effective potential with the radial coordinate in natural units for various values of cosmological constant $\Lambda$ in Kerr geometry with $a=0.99$.}
    \label{fig:vefker}
\end{figure}

Now, as we already did in Sec. \ref{subsec-1}, we  determine the effective potential using constants of motion. In fact, the effective potential turns out to be 
\begin{eqnarray}
V_{eff}=\frac{\mathcal{B}_r}{r^2}+P^2\nonumber \\-\frac{\chi^2}{r^4}\left[\left((r^2+a^2)P-a \Phi\right)^2  -\mathcal{B}_r(a P-\Phi)^2\right]
\end{eqnarray}
where, in the $a\rightarrow 0$ limit, we obtain the Schwarzchild-SG potential. We depict this effective potential in Fig. (\ref{fig:vefker}) for various values of $\Lambda$. Needless to say, circular orbits occur at $V'_{eff}=0$ and $V''_{eff}>0$. Then, following \cite{Slany:2020jhs}, requisite angular momenta for maintaining a circular orbit are found to be
\begin{eqnarray}
   \Phi_o=- \frac{2 a+ \frac{a r \mu^2 \Lambda}{12} (4 \frac{r^2}{\mu^2} + a^2) -r \left(a^2+\frac{4 r^2}{\mu ^2}\right) \sqrt{\frac{\mu }{2 r^3}-\frac{\Lambda }{3}} }{r \sqrt{1-\frac{1}{12} a^2 \Lambda  \mu ^2+a \mu  \sqrt{\frac{\mu }{2 r^3}-\frac{\Lambda }{3}}-\frac{3 \mu }{2 r}}} \nonumber \\ \label{phimax}
\end{eqnarray}
Using this formula, we can determine maximum and minimum angular momenta which form a circular orbit around a Kerr black hole in the SG framework. Taking the first derivative and solving for $\Lambda$ yields the $c_o$ values leading to real-valued circular orbits. From this analysis we find that $\Lambda$ can take values from zero to $0.14 \times (1/G^2_N M^2)$ and these two values  correspond, respectively, to the limit values $0.284(G_N M^2)<c_o<\infty$. There is no bound on negative values of the loop factor $c_o$ (corresponding to cases where number of bosonic degrees of freedom in the underlying QFT is larger than the bosonic ones). In Table \ref{tab:riscokerr}, tabulated are the innermost and outermost stable circular orbits in SG-dS and SG-AdS for various values of $c_o$ in Kerr geometry.
\begin{table}[!ht]
    \centering
    \begin{tabular}{ p{1.5cm} p{1cm} p{1cm} p{1.5cm} p{2cm} }
        \hline
        $\Lambda(1/\mu^2)$ & $r_{isco}(\mu)$ & $r_{osco} (\mu)$ & $c_o(G_N M^2)$ & $\Tilde{c}_o(CygX1)$  \\
        \hline
        \hline
        $10^{-2}$ & $0.77$ & $1,89$& \; $16$&  $2.7\times 10^{79}$\\
        $10^{-3}$ &  $0.73$ & $4.31$&  $160$&  $2.7\times 10^{80}$\\
        $10^{-6}$ &  $0.72$ & $45.1$&    $1.6\times10^{5}$&  $2.7\times 10^{83}$\\
        $-10^{-2}$ &  $0.68$ & $\;-$&    $-16$&  $-2.7\times 10^{79}$\\
        $-10^{-3}$ &  $0.72$ & $\;-$&    $-160$&  $-2.7\times 10^{80}$\\
        \hline
    \end{tabular}

    \caption{Stable orbits for symmergent dS-AdS Kerr black holes for various values of the loop factor $c_o$. Here, $\mu=2GM$ and $\Tilde{c_o}$ correspond to the X-Ray source $Cyg X-1$ with $M_{CygX1}=1.6 \times 10^{67} eV/c^2$. }
    \label{tab:riscokerr}
\end{table}

\section{Thin Accretion Disk onto   black hole in Symmergent Gravity}

In this section, we study thin accretion disk on  a black hole in symmergent gravity. To recall, one first notes tatt Shakura and Sunyaev proposed the standard model of geometrically thin accretion disks in 1973 using a Newtonian approach \cite{Shakura:1972te}. Next, this approach was generalized to general relativity \cite{novikov, Page:1974he} by  Novikov, Thorne and Page. Following these, thin accretion disks have been studied in modified theories of gravity for various spacetime geometries \cite{Harko:2009xf, Torres:2002td, Perez:2012bx, Soroushfar:2020kgb, Pun:2008ae, Staykov:2016dzn}.

The orbiting particles with the four-velocity $u^{\mu}$ form a disk of mean surface density $\Sigma$, where the rest mass density $\rho_{0}$, the energy flow vector $q^{\mu}$, and the stress tensor $t^{\mu\nu}$ are measured in the averaged rest-frame. Under these conditions, the surface density is given by 
\begin{equation}
\Sigma (r)=\int_{-H}^{H}\langle \rho_{0}\rangle dz,
\end{equation}%
in which $\langle \rho _{0}\rangle $ is the rest mass density averaged over time $\Delta t$ and angular range $2\pi$. Here, $H$ is the disk thickness, assumed to be much smaller than the disk radius.  The torque density can be written as
\begin{equation}
W_{\phi }{}^{r}=\int_{-H}^{H}\langle t_{\phi }{}^{r}\rangle dz,
\end{equation}%
where $\langle t_{\phi }^{r}\rangle$ is averaged again over the ranges $\Delta t$ and $2\pi$. The time and orbital averages of the energy flow vector $q^{\mu}$ gives the radiation flux ${\mathcal{F}}(r)$ over the disk surface as
\begin{equation}
{\mathcal{F}}(r)=\langle q^{z}\rangle .
\end{equation}
For the stress-energy tensor $T_{\nu}^{\mu}$ of the disk, the energy and angular momentum four-vectors are defined by $-E^{\mu}\equiv T_{\nu}^{\mu}(\partial/\partial t)^{\nu}$ and $J^{\mu}\equiv T_{\nu}^{\mu}(\partial/\partial\phi)^{\nu}$,  respectively. (Here the derivatives $\partial/\partial t$ and $\partial/\partial\phi$ stand actually for the Killing vectors.) 
The structure equations of the thin disk can be derived by using the conservations of the rest mass, energy, and the angular momentum \cite{novikov, Page:1974he}. In fact, from the rest mass conservation, $\nabla_{\mu}(\rho_{0}u^{\mu})=0$, it follows that the average rate of the accretion is independent of the disk radius,
\begin{equation}
\dot{M_{0}}\equiv -2\pi r\Sigma u^{r}=\mathrm{constant}.
\end{equation}
Then, in the steady-state approximation, specific energy $\widetilde{E}$ and specific angular momentum $\widetilde{L}$ of the accreting particles depend only on the radius of the orbits. Consequently, defining the black hole rotational velocity $\Omega =d\phi /dt$, energy conservation law $\nabla_{\mu}E^{\mu}=0$ leads to the motion integral
\begin{equation}
\lbrack \dot{M}_{0}\widetilde{E}-2\pi r\Omega W_{\phi }{}^{r}]_{,r}=4\pi r{\mathcal{F}}\widetilde{E}.\label{44}
\end{equation}%
This is a balance equation, which states that the energy transported by the rest mass flow, $\dot{M}_{0}\widetilde{E}$, and the energy transported by the torque in the disk, $2\pi r\Omega W_{\phi}^{r}$, are balanced by the energy radiated away from the surface of the disk, $4\pi r{\mathcal{F}}\widetilde{E}$.

On the other hand, angular momentum conservation, $\nabla_{\mu}J^{\mu}=0$, states the balance of three forms of angular momentum transport, viz.
\begin{equation}
\lbrack \dot{M}_{0}\widetilde{L}-2\pi rW_{\phi }{}^{r}]_{,r}=4\pi r{\mathcal{F}}\widetilde{L}\;. \label{45}
\end{equation}%
Now, eliminating $W_{\phi}^{r}$ from equations (\ref{44}) and (\ref{45}) and applying the energy-angular momentum relation for circular geodesic orbits in the form $\widetilde{E}_{,r}=\Omega \widetilde{L}_{,r}$, the flux $Q$ of the radiant energy (power) over the disk can be expressed as \cite{novikov, Page:1974he},
\begin{equation}
{\mathcal{F}}\left(r\right) =-\frac{\dot{M_{0}}}{4\pi\sqrt{-g}}\frac{\Omega _{,r}}{\left( \widetilde{E}-\Omega \widetilde{L}\right) ^{2}}\int_{r_{\textmd{\scriptsize{ms}}}}^{r}\left( \widetilde{E}-\Omega \widetilde{L}\right) \widetilde{L}_{,r}dr\, .\label{flux}
\end{equation}%
The disk is supposed to be in thermodynamical equilibrium so the radiation flux emitted by the disk surface will follow Stefan-Boltzmann law:%
\begin{equation}
{\mathcal{F}}\left(r\right) =\sigma T^{4}\left( r\right)\, ,
\label{stefan}
\end{equation}%
where $\sigma $ is the Stefan-Boltzmann constant. Since flux is a local quantity that a faraway observer cannot observe, one can also define luminosity $L_\infty$ (energy per unit time) that reaches an observer at infinity. It is obtained from the relation 
\begin{eqnarray}
    \frac{dL_\infty}{d \log r}=4 \pi r \sqrt{-g} \widetilde{E} {\mathcal{F}}\, .
    \label{diff-lumi}
\end{eqnarray}
Involving the conserved quantities, the effective potential takes the form
\begin{equation}
V_{\textmd{\scriptsize{eff}}}(r) = -1 + \frac{\widetilde{E}^{2}g_{\phi\phi} + 2\widetilde{E}\widetilde{L}g_{t\phi} - \widetilde{L}^{2}g_{tt}}{g_{t\phi}^{2} + g_{tt}g_{\phi\phi}}\,.
\label{Veff-x}
\end{equation}%
Then, existence of circular orbits at any arbitrary radius $r$ in the equatorial plane demands that $V_{\textmd{\scriptsize{eff}}}(r)=0$ and $dV_{\textmd{\scriptsize{eff}}}/dr=0$. These conditions allow us to write the kinematic parameters in (\ref{Veff-x}) as \cite{Harko:2009xf}
\begin{eqnarray}
\widetilde{E} &=&\frac{g_{tt}-g_{t\phi}\Omega}{\sqrt{g_{tt}-2g_{t\phi}\Omega - g_{\phi\phi}\Omega^{2}}}\,, \\
\widetilde{L} &=&\frac{g_{t\phi}+g_{\phi\phi}\Omega}{\sqrt{g_{tt} - 2g_{t\phi}\Omega - g_{\phi\phi}\Omega^{2}}}\,, \\
\Omega &=&=\frac{-g_{t\phi ,r}+\sqrt{(g_{t\phi,r})^{2}+g_{tt,r}g_{\phi \phi ,r}}}{g_{\phi \phi ,r}}\,,
\end{eqnarray}%
in which $X_{,r}\equiv dX/dr$ throughout the analysis. Stability of orbits depends on the signs of $d^{2}V_{\textmd{\scriptsize{eff}}}/dr^{2}$, while the condition $d^{2}V_{\textmd{\scriptsize{eff}}}/dr^{2}=0$ gives the inflection point or \textit{marginally stable} (ms) orbit or innermost stable circular orbit (isco) at $r=r_{\textmd{\scriptsize{ms}}}$. The thin disk is assumed to have an inner edge defined by the marginally stable circular radius $r_{\textmd{\scriptsize{ms}}}$, while the orbits at radii larger than $r=r_{\textmd{\scriptsize{ms}}}$ are Keplerian. One notes that marginally stable circular orbits become crucial when accretion surrounds the black hole.

Another essential characteristic of the thin accretion disk is its efficiency $\epsilon$, which quantifies the central body's ability to convert the accreting mass into radiation. The efficiency is measured at infinity, defined as the ratio of two rates: the rate of energy of the photons emitted from the disk surface and the rate with which the mass energy is transported to the central body. If all photons reach infinity, the Page-Thorne accretion efficiency is given by the specific energy of the accreting particles measured at the marginally stable orbit \cite{Page:1974he}:
\begin{equation}
\epsilon =1-\widetilde{E}\left(r_{\textmd{\scriptsize{ms}}}\right),
\end{equation}%
As the definition indicates, $\epsilon$ should be non-negative.

In the two subsections below, We will study static and stationary cases. In fact, we will adopt a black hole mass $M_{CygX1}=1.6 \times 10^{67} eV/c^2$ and accretion rate $\dot{M_{0}}=4.7 \times 10^{18}g s^{-1}$ \cite{Gou:2011nq}. We will consider a maximally rotating ($a=0.99$) black hole for the stationary case. We will take the outer edge of the ring as $r_{out}=11 r_{isco}$ when necessary \cite{Dove:1997ei}.

\subsection{Schwarzchild Case}
\begin{figure}[!ht]
    \centering
    \includegraphics[scale=0.6]{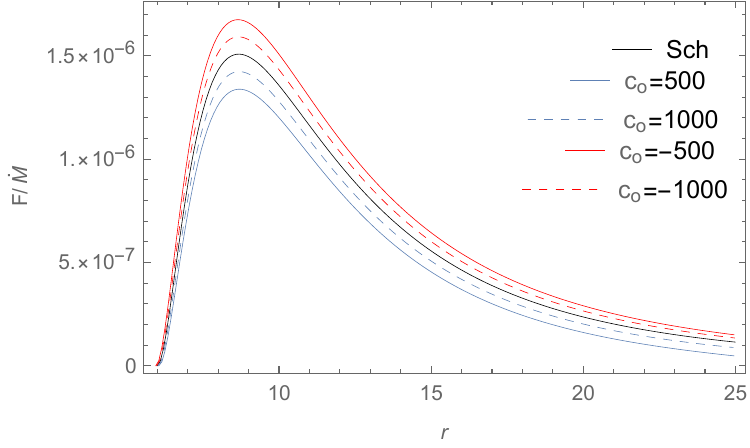}
    \caption{Variation of the energy flux with the radial coordinate $r$ in natural units for various values of $c_o$ in static case for $a=0.99$.}
    \label{fig:fsch}
\end{figure}

\begin{figure}[!ht]
    \centering
    \includegraphics[scale=0.6]{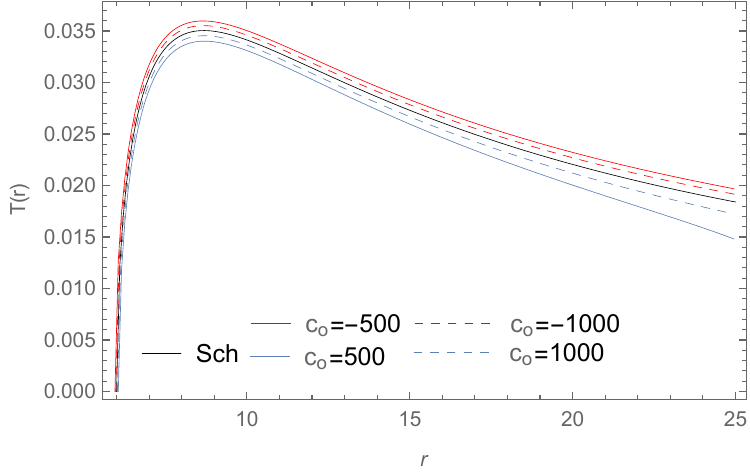}
    \caption{Variation of the temperature with the radial coordinate $r$ in natural units for various values of $c_o$ with $a=0.99$ in static case.}
    \label{fig:tsch}
\end{figure}

\begin{figure}[!ht]
    \centering
    \includegraphics[scale=0.6]{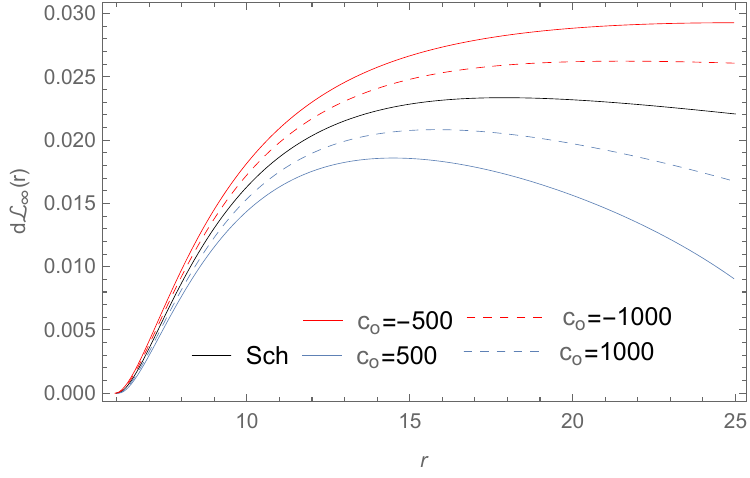}
    \caption{Variation of the differential luminosity with the radial coordinate $r$  in natural units for various values of $c_o$ with $a=0.99$ in the static case.}
    \label{fig:lsch}
\end{figure}

For the static metric (\ref{static}),  kinematic parameters in (\ref{Veff-x}) take the values 
\begin{eqnarray}
    \Omega&=&\frac{r \sqrt{\frac{3 \mu }{r}-2 \Lambda  r^2}}{\sqrt{6-\frac{9 \mu }{r}}}\\
    \widetilde{E}&=&\frac{\sqrt{\frac{3 \mu }{r}-2 \Lambda  r^2}}{\sqrt{6} r} \\
    \widetilde{L}&=&-\frac{3 \mu +\Lambda  r^3-3 r}{3 r \sqrt{1-\frac{3 \mu }{2 r}}} 
\end{eqnarray}
Given these relations, one can then calculate the energy flux from (\ref{flux}) for a particular value of $\Lambda$ by first determining $r_{isco}$ from $V_{eff}$ at that particular value. We numerically integrate the flux (\ref{flux}) and calculate temperature $T$ via Stefan Boltzmann law in (\ref{Veff-x}) and the differential luminosity $L_\infty$ in (\ref{diff-lumi}) for various values of loop factor $c_o$ and plot them in Figs. \ref{fig:fsch}, \ref{fig:tsch} and \ref{fig:lsch}. We tabulate in Table  table \ref{tab:numsc} our numerical calculations of the efficiency values and the maximum temperature of the disk. As $c_o$ approaches to infinity from both the positive and negative side, the values of ${\mathcal{F}}$, $T$ and $L_\infty$ all approach to the Schwarzchild value asymptotically. As $c_o$ approaches to zero from positive (negative) values, the net flux, temperature and luminosity decrease (increase). This is due to inverse correlation between $c_o$ and $\Lambda$ as because $\Lambda \approx 1/c_o$. We see that the loop factor $c_o$ does not significantly differ from the Schwarzchild case considering its required high value.
\begin{table}[!hbt]
    \centering
    \begin{tabular}{p{1.5cm} p{2cm} p{2cm} p{1.5cm}}
        \hline
         $c_o(G_N M^2)$& $\Tilde{c}_o(CygX1)$ & $F_{Max}(keV^4)$ & $T_{Max}(keV)$ \\
        \hline
        \hline
        $0$ & $0$ &$2.28\times10^{-4}$ & $0.192$ \\
        $500$ & $9\times10^{80}$ & $2.02 \times 10^{-4}$ & $0.187$ \\
        $1000$ & $1.8\times10^{81}$ & $2.15 \times 10^{-4}$ & $0.190$ \\
        $-500$ & $-9\times10^{80}$ & $2.53 \times 10^{-4}$ & $0.198$ \\
        $-1000$ & $-1.8\times 10^{81}$ & $2.40 \times 10^{-4}$ & $0.195$ \\
        \hline
        
    \end{tabular}

    \caption{Maximum flux and temperature values for symmergent dS-AdS Schwarzchild black holes for various $c_o$ values for the X-Ray source $Cyg X-1$ with mass $M_{CygX1}=1.6 \times 10^{67} eV/c^2$. }
    \label{tab:numsc}
\end{table}


\subsection{Kerr case}
\begin{figure}[!hbt]
    \centering
    \includegraphics[scale=0.6]{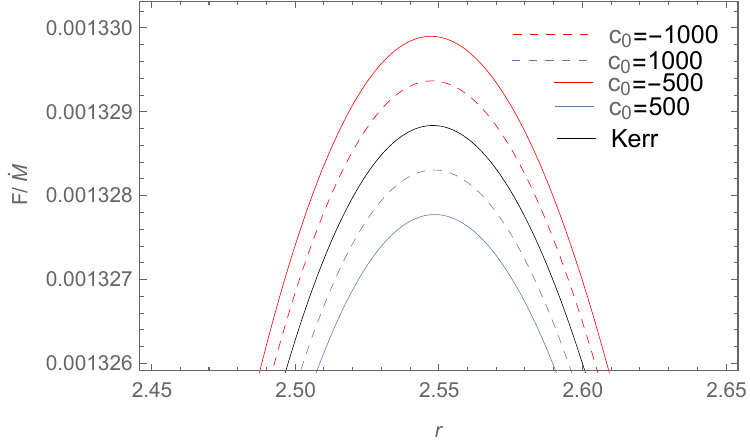}
    \caption{Variation of the energy flux with the radial coordinate in natural units for various values of the loop factor $c_o$ for Kerr black hole with $a=0.99$.}
    \label{fig:fkerr}
\end{figure}

\begin{figure}[!hbt]
    \centering
    \includegraphics[scale=0.6]{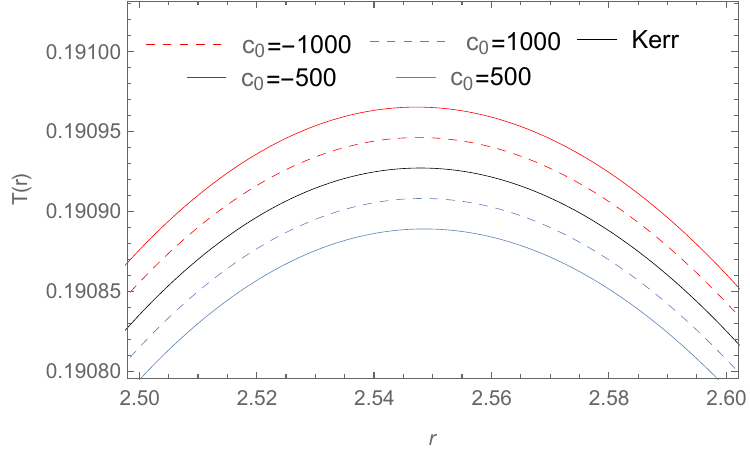}
    \caption{Variation of temperature with the radial coordinate in natural units for various values of the loop factor $c_o$ for $a=0.99$ in the stationary case.}
    \label{fig:tkerr}
\end{figure}

\begin{figure}[!hbt]
    \centering
    \includegraphics[scale=0.6]{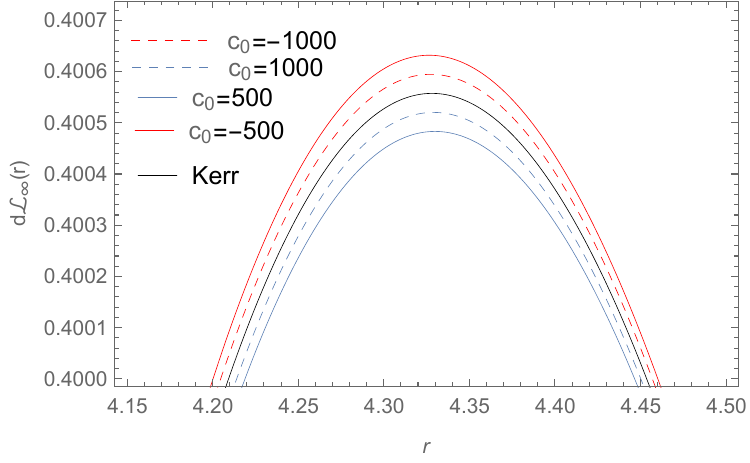}
    \caption{Variation of differential luminosity with the radial coordinate in natural units for various values of the loop factor $c_o$ for $a=0.99$ in the stationary case.}
    \label{fig:lkerr}
\end{figure}

In this subsection, we calculate the kinetic parameters in (\ref{Veff-x}) for the stationary metric in (\ref{equatorialmetric}). In this regard, the kinetic parameters turn out to be 
\begin{eqnarray}
    \Omega&=&\frac{k_1}{2 r^3 \left(a^2 \Lambda +3\right)-3 a^2 \mu }\\
     \widetilde{E}&=&\frac{k_2+k_3 k_0}{k_4 \sqrt{\frac{k_5+k_6}{k_7}}} \\
     \widetilde{L}&=&\frac{k_8+k_9}{k_4 \sqrt{\frac{k_5+k_6}{k_7}}}
\end{eqnarray}
in which the individual parameters $k_i$ are given by
\begin{eqnarray}
    k_0=\sqrt{\frac{\left(2 \Lambda  r^3-3 \mu \right) \left(2 r^3 \left(a^2 \Lambda -1\right)-3 a^2 \mu \right)}{r^4 \left(a^2 \Lambda +3\right)^4}}
\end{eqnarray}
\begin{eqnarray}
    k_1=6 \sqrt{3} a^2 \Lambda  r^2 k_0+9 \sqrt{3} r^2 k_0+\sqrt{3} a^4 \Lambda ^2 r^2 k_0-6 a \mu\nonumber \\+4 a \Lambda  r^3
\end{eqnarray}
\begin{eqnarray}
    k_2=6 r^4 \left(a^4 \Lambda ^2+3\right)-27 a^2 \mu ^2+6 \Lambda  r^6 \left(a^2 \Lambda -1\right) \nonumber \\ -9 a^2 \mu  r \left(a^2 \Lambda +1\right)+9 r^3 \mu  \left(a^2 \Lambda -2\right) 
\end{eqnarray}
\begin{eqnarray}
    k_3=2 \sqrt{3} a \Lambda  r^5 \left(a^2 \Lambda +3\right)^2 +6 \sqrt{3} a \mu  r^2 \left(a^2 \Lambda +3\right)^2\nonumber \\+ 2 \sqrt{3} a^3 r^3 \Lambda  \left(a^2 \Lambda +3\right)^2 
\end{eqnarray}
\begin{eqnarray}  
    k_4=\left(a^2 \Lambda +3\right)^2 \left(2 r^4 \left(a^2 \Lambda +3\right)-3 a^2 \mu  r\right)
\end{eqnarray}
\begin{eqnarray}    
    k_5=-54 a^4 \mu ^2+4 r^6 \left(a^2 \Lambda +3\right)^2+18 \mu  r^5 \left(7 a^2 \Lambda -3\right) \nonumber \\+6 a^2 \mu  r^3 \left(5 a^2 \Lambda -9\right)
\end{eqnarray}
\begin{eqnarray}
    k_6=36 \sqrt{3} a \mu  r^4 \left(a^2 \Lambda +3\right)^2 k_0 \nonumber \\+3 a^2 \mu  r^2 \left(4 \sqrt{3} a \left(a^2 \Lambda +3\right)^2 k_0-63 \mu \right)
\end{eqnarray}
\begin{eqnarray}   
    k_7=\left(a^2 \Lambda +3\right)^2 \left(2 r^3 \left(a^2 \Lambda +3\right)-3 a^2 \mu \right)^2
\end{eqnarray}
\begin{eqnarray}    
    k_8=-18 a^3 \mu +27 \sqrt{3} a^2 r \left(\mu +\Lambda  r^3+r\right) k_0 \nonumber \\ +27 \sqrt{3} r^4 k_0+\sqrt{3} a^8 \Lambda ^3 r^2 k_0 
\end{eqnarray}
\begin{eqnarray}    
    k_9=\sqrt{3} a^6 \Lambda ^2 r \left(3 \mu +\Lambda  r^3+9 r\right) k_0 \nonumber \\ +9 \sqrt{3} a^4 \Lambda  r \left(2 \mu +\Lambda  r^3+3 r\right) k_0-54 a \mu  r^2
\end{eqnarray}
Given these parameters, we numerically integrate the flux in equation \ref{flux}) for any particular value of $\Lambda$ by finding the corresponding $r_{isco}$ at that particular value. In Figs.  \ref{fig:fkerr}, \ref{fig:tkerr} and \ref{fig:lkerr}, we plot flux, temperature and differential luminosity for various $c_o$ values. Also, in the Table \ref{tab:numkerr}, we numerically calculate maximum temperature and efficiency values for different $c_o$ values. In general, we do not observe any significant deviation from the standard Kerr limit due to the expected high value of the loop factor $c_o$. 
\begin{table}[!hbt]
    \centering
    \begin{tabular}{p{1.5cm} p{2cm} p{2cm} p{1.5cm}}
        \hline
         $c_o(G_N M^2)$& $\Tilde{c}_o(CygX1)$ & $F_{Max}(keV^4)$ & $T_{Max}(keV)$ \\
        \hline
        \hline
        $0$ & $0$ &$0.20188$ & $1.05155$ \\
        $500$ & $9\times10^{80}$ & $0.200886$ & $1.05124 $ \\
        $1000$ & $1.8\times10^{81}$ & $0.201135$ & $1.05156$ \\
        $-500$ & $-9\times10^{80}$ & $0.201366$ & $1.05186$ \\
        $-1000$ & $-1.8\times 10^{81}$ & $0.201116 $ & $1.05154 $ \\
        \hline
        
    \end{tabular}

    \caption{Maximum flux and temperature values for symmergent dS-AdS Kerr black holes with various loop factors $c_o$ for the X-Ray source $Cyg X-1$ with $M_{CygX1}=1.6 \times 10^{67} eV/c^2$. }
    \label{tab:numkerr}
\end{table}
\section{Discussion}
In this section, we give a critical discussion of the various results we found in the two sections above. Basically, we found four distinct determinations on the loop factor $c_o$ from the circular orbits in symergent gravity. The first bound  is $c_{o}^{(1)}=0.284\times (G_N M^2)$ such that  beyond this value no naked singularities occur in the static case. Second, is $c_{o}^{(2)}=0.358\times (G_N M^2)$ such that beyond this value circular orbits can form in stationary solutions. The third is  $c_{o}^{(3)}=55.95\times (G_N M^2)$ such that beyond this value circular orbits can also occur in Schwarzchild black holes in the SG framework. For the negative loop factor, however, we have only one determination $c_{o}^{(4)}=-1.32\times (G_N M^2)$ and, below this limit, no naked singularities occur. In fact, in Table \ref{tab:desctab} we give value of the loop factor for a given black hole with $M_{CygX1}$.
\begin{table}[!hbt]
    \centering
    \begin{tabular}{p{1.5cm} p{2cm} p{3.cm}}
        \hline
         $c_o(G_N M^2)$& $\Tilde{c}_o(CygX1)$ & Description of the Bound \\
        \hline
        \hline
        $-\infty$ & $-\infty$ &GR limit  \\
        \hline
        $55.95$ & $9.60\times 10^{79}$ & Circular orbit can form for Schwarzchild blackholes \\
        \hline
        $0.358$ & $6.14\times 10^{77}$ & Circular orbit can form for Kerr blackholes \\
        \hline
        $0.284$ & $4.82\times 10^{77}$ & No naked singularities for Schwarzchild blackholes \\
        \hline
        $0$ & $0$ & - \\
        \hline
        $-1.32$ & $-2.26\times 10^{78}$ & No naked singularities for Kerr blackholes \\
        \hline
        $-\infty$ & $-\infty$ & GR limit \\
        \hline
        \hline
        
    \end{tabular}

     \caption{Bounds on SG parameters. The table describes bounds on the loop factor as $c_o$ increases (decreases) from 0 when $c_o>0$ ($c_o<0$). Here, $\Tilde{c_o}$ value is the exact value of the loop coefficients for the X-Ray source $Cyg X-1$ with $M_{CygX1}=1.6 \times 10^{67} eV/c^2$. One notices that the loop factor $c_o$ is inversely proportional to $\Lambda$ ($c_o \propto 1/\Lambda$).}
    \label{tab:desctab}
\end{table}
We also find that positive (negative) loop factors cause temperature and luminosity of the disk to decrease (increase).   As $c_o$ increases and asymptotically approaches to infinity from both positive and negative sides, the temperature and luminosity profiles reduce to the GR values. Regarding efficiency of both Schwarzchild and Kerr cases we find nearly no deviation from the original values (about \%6 for Schwarzchild and \%42 for Kerr case).

\section{Conclusion}
Here we conclude our analysis of the symmergent gravity -- a recent proposal offering an emergent gravity framework starting from flat spacetime effective field theory by way of gauge symmetry restoration. 
Symmergent gravity allows one to probe or constrain field-theoretic (or elementary particle) properties in strong gravity media like black holes, stellar objects and the early Universe. In this paper, we focused on investigation of circular orbits and accretion disk onto the symmergent black holes. 

In Sec. III, we investigated the circular orbits of the symmergent black holes in Schwarzschild and Kerr geometries. The loop factor $c_o$ directly affects circular orbits' effective potential and radius. In Figs. \ref{fig:horizonsc} and \ref{fig:horizonkerr}, for instance, $c_o$ determines the horizon number through the effective cosmological constant $\Lambda$. For symmergent gravity with heavy fermionic (bosonic) fields, we have a dS (AdS) type black holes in Figs. \ref{fig:veffsch} and \ref{fig:vefker}. We also see that $c_o$ directly determines the characteristics of the effective potential. We find numerous bounds on $c_o$ by examining circular orbits (Table \ref{tab:desctab}). In fact, we calculated several radii  of the circular orbits for different values of the  loop coefficients in tables \ref{tab:risco} and \ref{tab:riscokerr}.

In section IV, we have calculated the radiated flux, temperature and differential luminosities of Schwarzchild and Kerr black holes in symmergent gravity framework. We tabulated our findings in Tables \ref{tab:numsc} and \ref{tab:numkerr}, and plotted in Figs. \ref{fig:fsch},\ref{fig:tsch},\ref{fig:lsch},\ref{fig:fkerr}, \ref{fig:tkerr} and \ref{fig:lkerr}. 

Since the entire effect of the loop factor $c_o$ acts as the cosmological constant, it directly affects black hole physics. In fact, $c_o$ is directly tied to the difference between the number of bosons and fermions in the underlying QFT. Also, symmergent gravity is an emergent gravity theory and quadratic-curvature term already contains the Planck mass ($m_p = (8\pi\sqrt{G_N})^{-1/2}$) besides the loop factor $c_o$. Thus, to compensate for the $m_p$, number of bosonic and fermionic must be significantly large.   This is a direct indication that new particles exist beyond the known ones forming up the observable Universe. The new particles, which could be small or large compared to some $10^{80}$ hydrogen atoms forming up the observable Universe, do actually not have to interact with the known particles. They can actually form a dark sector where we can extract information via only their gravitational effects -- as we do here using the black hole environments. These dark sector particles form a naturally-coupled sector because their naturally feeble couplings agree with all the existing bounds on dark matter and dark energy (see the review volume \cite{demir-mdpi} as well as \cite{demir-cankocak}). 

\acknowledgments 
A.{\"O}. would like to acknowledge the contribution of the COST Action CA18108 - Quantum gravity phenomenology in the multi-messenger approach (QG-MM). We also thank the anonymous referee for a constructive report.


\begin{thebibliography}{99}

\bibitem{EventHorizonTelescope:2019dse}
K.~Akiyama \textit{et al.} [Event Horizon Telescope],
``First M87 Event Horizon Telescope Results. I. The Shadow of the Supermassive black hole,''
Astrophys. J. Lett. \textbf{875}, L1 (2019).

\bibitem{EventHorizonTelescope:2019ggy}
K.~Akiyama \textit{et al.} [Event Horizon Telescope],
``First M87 Event Horizon Telescope Results. VI. The Shadow and Mass of the Central black hole,''
Astrophys. J. Lett. \textbf{875}, no.1, L6 (2019).



\bibitem{Akiyama:2022tyh}
K.~Akiyama, A.~Alberdi, W.~Alef, J.~C.~Algaba, R.~Anantua, K.~Asada, R.~Azulay, U.~Bach, A.~K.~Baczko and D.~Ball, \textit{et al.}
``First Sagittarius A* Event Horizon Telescope Results. I. The Shadow of the Supermassive black hole in the Center of the Milky Way,''
Astrophys. J. Lett. \textbf{930}, no.2, L12 (2022).




\bibitem{Akiyama:2022yot}
K.~Akiyama, A.~Alberdi, W.~Alef, J.~Carlos Algaba, R.~Anantua, K.~Asada, R.~Azulay, U.~Bach, A.~K.~Baczko and D.~Ball, \textit{et al.}
``First Sagittarius A* Event Horizon Telescope Results. V. Testing Astrophysical Models of the Galactic Center black hole,''
Astrophys. J. Lett. \textbf{930}, no.2, L16 (2022).

\bibitem{Akiyama:2022qhc}
K.~Akiyama, A.~Alberdi, W.~Alef, J.~C.~Algaba, R.~Anantua, K.~Asada, R.~Azulay, U.~Bach, A.~K.~Baczko and D.~Ball, \textit{et al.}
``First Sagittarius A* Event Horizon Telescope Results. VI. Testing the black hole Metric,''
Astrophys. J. Lett. \textbf{930}, no.2, L17 (2022).




\bibitem{dyson}
F.~Dyson,
``Is a graviton detectable?,''
Int. J. Mod. Phys. A \textbf{28}, 1330041 (2013).

\bibitem{thooft}
G.~'t Hooft and M.~J.~G.~Veltman,
``One loop divergencies in the theory of gravitation,''
Ann. Inst. H. Poincare Phys. Theor. A \textbf{20} (1974), 69-94.



\bibitem{incompatible}
A.~Macias and A.~Camacho,
``On the incompatibility between quantum theory and general relativity,''
Phys. Lett. B \textbf{663}, 99-102 (2008).


\bibitem{weinberg}
S.~Weinberg,
``Phenomenological Lagrangians,''
Physica A \textbf{96}, no.1-2, 327-340 (1979).


\bibitem{eft}
I.~Brivio and M.~Trott,
``The Standard Model as an Effective Field Theory,''
Phys. Rept. \textbf{793}, 1-98 (2019).


\bibitem{sakharov}
A.~D.~Sakharov,
``Vacuum quantum fluctuations in curved space and the theory of gravitation,''
Usp. Fiz. Nauk \textbf{161}, no.5, 64-66 (1991).


\bibitem{visser}
M.~Visser,
``Sakharov's induced gravity: A Modern perspective,''
Mod. Phys. Lett. A \textbf{17}, 977-992 (2002).




\bibitem{verlinde}
E.~P.~Verlinde,
``Emergent Gravity and the Dark Universe,''
SciPost Phys. \textbf{2} (2017) no.3, 016.




\bibitem{demir1}
D.~Demir,
``Emergent Gravity as the Eraser of Anomalous Gauge Boson Masses, and QFT-GR Concord,''
Gen. Rel. Grav. \textbf{53}, no.2, 22 (2021).




\bibitem{demir2}
D.~Demir,
``Symmergent Gravity, Seesawic New Physics, and their Experimental Signatures,''
Adv. High Energy Phys. \textbf{2019}, 4652048 (2019).


\bibitem{demir3}
D.~A.~Demir,
``Curvature-Restored Gauge Invariance and Ultraviolet Naturalness,''
Adv. High Energy Phys. \textbf{2016}, 6727805 (2016).


\bibitem{irfan}
{\.I}.~{\.I}.~{\c C}imdiker,
``Starobinsky inflation in emergent gravity,''
Phys. Dark Univ. \textbf{30}, 100736 (2020).


\bibitem{Cimdiker:2021cpz}
\.I.~\c{C}imdiker, D.~Demir and A.~\"Ovg\"un,
``Black hole shadow in symmergent gravity,''
Phys. Dark Univ. \textbf{34}, 100900 (2021).



\bibitem{birrel}
N. Birrell and P. Davies, {\it Quantum Fields in Curved Space} (Cambridge University Press, Cambridge, 1982). 



\bibitem{f(R)review1}
T.~P.~Sotiriou and V.~Faraoni,
``f(R) Theories Of Gravity,''
Rev. Mod. Phys. \textbf{82}, 451-497 (2010).

\bibitem{Clifton:2011jh}
T.~Clifton, P.~G.~Ferreira, A.~Padilla and C.~Skordis,
``Modified Gravity and Cosmology,''
Phys. Rept. \textbf{513}, 1-189 (2012).

\bibitem{DeFelice:2010aj}
A.~De Felice and S.~Tsujikawa,
``f(R) theories,''
Living Rev. Rel. \textbf{13}, 3 (2010).

\bibitem{Olmo:2011uz}
G.~J.~Olmo,
``Palatini Approach to Modified Gravity: f(R) Theories and Beyond,''
Int. J. Mod. Phys. D \textbf{20}, 413-462 (2011).


\bibitem{Cembranos:2011sr}
J.~A.~R.~Cembranos, A.~de la Cruz-Dombriz and P.~Jimeno Romero,
``Kerr-Newman black holes in $f(R)$ theories,''
Int. J. Geom. Meth. Mod. Phys. \textbf{11}, 1450001 (2014).


\bibitem{Nashed:2019tuk}
G.~G.~L.~Nashed and S.~Capozziello,
``Charged spherically symmetric black holes in $f(R)$ gravity and their stability analysis,''
Phys. Rev. D \textbf{99}, no.10, 104018 (2019).

\bibitem{Elizalde:2020icc}
E.~Elizalde, G.~G.~L.~Nashed, S.~Nojiri and S.~D.~Odintsov,
``Spherically symmetric black holes with electric and magnetic charge in extended gravity: physical properties, causal structure, and stability analysis in Einstein\textquoteright{}s and Jordan\textquoteright{}s frames,''
Eur. Phys. J. C \textbf{80}, no.2, 109 (2020).

\bibitem{Nashed:2021sey}
G.~G.~L.~Nashed,
``New rotating AdS/dS black holes in $f(R)$ gravity,''
Phys. Lett. B \textbf{815}, 136133 (2021).

\bibitem{Zheng:2018fyn}
Y.~Zheng and R.~J.~Yang,
``Horizon thermodynamics in $f(R)$ theory,''
Eur. Phys. J. C \textbf{78}, no.8, 682 (2018).


\bibitem{Moon:2011sz}
T.~Moon, Y.~S.~Myung and E.~J.~Son,
``Stability analysis of f(R)-AdS black holes,''
Eur. Phys. J. C \textbf{71}, 1777 (2011).

\bibitem{Myung:2011ih}
Y.~S.~Myung, T.~Moon and E.~J.~Son,
``Stability of f(R) black holes,''
Phys. Rev. D \textbf{83}, 124009 (2011).


\bibitem{Stuchlik:1999qk}
Z.~Stuchlik and S.~Hledik,
``Some properties of the Schwarzschild-de Sitter and Schwarzschild - anti-de Sitter space-times,''
Phys. Rev. D \textbf{60}, 044006 (1999).

\bibitem{Rezzolla:2003re}
L.~Rezzolla, O.~Zanotti and J.~A.~Font,
``Dynamics of thick discs around Schwarzschild-de Sitter black holes,''
Astron. Astrophys. \textbf{412}, 603 (2003).

\bibitem{Perlick:2015vta}
V.~Perlick, O.~Y.~Tsupko and G.~S.~Bisnovatyi-Kogan,
``Influence of a plasma on the shadow of a spherically symmetric black hole,''
Phys. Rev. D \textbf{92}, no.10, 104031 (2015).


\bibitem{Luminet:1979nyg}
J.~P.~Luminet,
``Image of a spherical black hole with thin accretion disk,''
Astron. Astrophys. \textbf{75}, 228-235 (1979).

\bibitem{Falcke:1999pj}
H.~Falcke, F.~Melia and E.~Agol,
``Viewing the shadow of the black hole at the galactic center,''
Astrophys. J. Lett. \textbf{528}, L13 (2000).




\bibitem{Ovgun:2018tua}
A.~\"Ovg\"un, \.I.~Sakall\i{} and J.~Saavedra, ``Shadow cast and Deflection angle of Kerr-Newman-Kasuya spacetime,''
JCAP \textbf{10}, 041 (2018).

\bibitem{Ovgun:2020gjz}
A.~\"Ovg\"un and \.I.~Sakall\i{},
``Testing generalized Einstein-Cartan-Kibble-Sciama gravity using weak deflection angle and shadow cast,''
Class. Quant. Grav. \textbf{37}, no.22, 225003 (2020).

\bibitem{Okyay:2021nnh}
M.~Okyay and A.~\"Ovg\"un,
``Nonlinear electrodynamics effects on the black hole shadow, deflection angle, quasinormal modes and greybody factors,''
JCAP \textbf{01}, no.01, 009 (2022).

\bibitem{Pantig:2021zqe}
R.~C.~Pantig, P.~K.~Yu, E.~T.~Rodulfo and A.~\"Ovg\"un,
``Shadow and weak deflection angle of extended uncertainty principle black hole surrounded with dark matter,''
Annals of Physics 436, 168722 (2022).

\bibitem{Pantig:2022whj}
R.~C.~Pantig and A.~\"Ovg\"un,
``Dehnen halo effect on a black hole in an ultra-faint dwarf galaxy,''
JCAP \textbf{08}, no.08, 056 (2022).

\bibitem{Pantig:2022toh}
R.~C.~Pantig and A.~\"Ovg\"un,
``Dark matter effect on the weak deflection angle by black holes at the center of Milky Way and M87 galaxies,''
Eur. Phys. J. C \textbf{82}, no.5, 391 (2022).


\bibitem{Pantig:2022qak}
R.~C.~Pantig, A.~\"Ovg\"un and D.~Demir,
``Testing Symmergent gravity through the shadow image and weak field photon deflection by a rotating black hole using the M87$^*$ and Sgr. A$^*$ results,''
Eur. Phys. J. C 83, 250 (2023).




\bibitem{Rayimbaev:2022hca}
J.~Rayimbaev, R.~C.~Pantig, A.~\"Ovg\"un, A.~Abdujabbarov and D.~Demir,
``Quasiperiodic oscillations, weak field lensing and shadow cast around black holes in Symmergent gravity,'', 	arXiv:2206.06599 [gr-qc],
Annals of Physics  (2023).

\bibitem{Pantig:2022ely}
R.~C.~Pantig and A.~\"Ovg\"un,
``Testing dynamical torsion effects on the charged black hole\textquoteright{}s shadow, deflection angle and greybody with M87* and Sgr. A* from EHT,''
Annals Phys. \textbf{448}, 169197 (2023).



\bibitem{Uniyal:2022vdu}
A.~Uniyal, R.~C.~Pantig and A.~\"Ovg\"un,
``Probing a nonlinear electrodynamics black hole with thin accretion disk, shadow and deflection angle with M87* and Sgr A* from EHT,''
Physics of the Dark Universe 40 (2023) 101178.

\bibitem{Gralla:2019xty}
S.~E.~Gralla, D.~E.~Holz and R.~M.~Wald,
``black hole Shadows, Photon Rings, and Lensing Rings,''
Phys. Rev. D \textbf{100}, no.2, 024018 (2019).




\bibitem{Takahashi:2004xh}
R.~Takahashi,
``Shapes and positions of black hole shadows in accretion disks and spin parameters of black holes,''
J. Korean Phys. Soc. \textbf{45}, S1808-S1812 (2004).







\bibitem{Allahyari:2019jqz}
A.~Allahyari, M.~Khodadi, S.~Vagnozzi and D.~F.~Mota,
``Magnetically charged black holes from non-linear electrodynamics and the Event Horizon Telescope,''
JCAP \textbf{02}, 003 (2020).





\bibitem{Dokuchaev:2020wqk}
V.~I.~Dokuchaev and N.~O.~Nazarova,
``Visible shapes of black holes M87* and SgrA*,''
Universe \textbf{6}, no.9, 154 (2020).



\bibitem{Zeng:2020dco}
X.~X.~Zeng, H.~Q.~Zhang and H.~Zhang,
``Shadows and photon spheres with spherical accretions in the four-dimensional Gauss\textendash{}Bonnet black hole,''
Eur. Phys. J. C \textbf{80}, no.9, 872 (2020).






\bibitem{Paul:2019trt}
S.~Paul, R.~Shaikh, P.~Banerjee and T.~Sarkar,
``Observational signatures of wormholes with thin accretion disks,''
JCAP \textbf{03}, 055 (2020).



\bibitem{Johannsen:2015mdd}
T.~Johannsen,
``Sgr A* and General Relativity,''
Class. Quant. Grav. \textbf{33}, no.11, 113001 (2016).




\bibitem{Johannsen:2016vqy}
T.~Johannsen, C.~Wang, A.~E.~Broderick, S.~S.~Doeleman, V.~L.~Fish, A.~Loeb and D.~Psaltis,
``Testing General Relativity with Accretion-Flow Imaging of Sgr A*,''
Phys. Rev. Lett. \textbf{117}, no.9, 091101 (2016).




\bibitem{Johannsen:2015hib}
T.~Johannsen, A.~E.~Broderick, P.~M.~Plewa, S.~Chatzopoulos, S.~S.~Doeleman, F.~Eisenhauer, V.~L.~Fish, R.~Genzel, O.~Gerhard and M.~D.~Johnson,
``Testing General Relativity with the Shadow Size of Sgr A*,''
Phys. Rev. Lett. \textbf{116}, no.3, 031101 (2016).

\bibitem{Broderick:2016ewk}
A.~E.~Broderick, V.~L.~Fish, M.~D.~Johnson, K.~Rosenfeld, C.~Wang, S.~S.~Doeleman, K.~Akiyama, T.~Johannsen and A.~L.~Roy,
``Modeling Seven Years of Event Horizon Telescope Observations with Radiatively Inefficient Accretion Flow Models,''
Astrophys. J. \textbf{820}, no.2, 137 (2016).


\bibitem{Rahaman:2021web}
F.~Rahaman, K.~N.~Singh, R.~Shaikh, T.~Manna and S.~Aktar,
``Shadows of Lorentzian traversable wormholes,''
Class. Quant. Grav. \textbf{38}, no.21, 215007 (2021).

\bibitem{Amir:2018szm}
M.~Amir, A.~Banerjee and S.~D.~Maharaj,
``Shadow of charged wormholes in Einstein\textendash{}Maxwell\textendash{}dilaton theory,''
Annals Phys. \textbf{400}, 198-207 (2019).


\bibitem{Shaikh:2021cvl}
R.~Shaikh, S.~Paul, P.~Banerjee and T.~Sarkar,
``Shadows and thin accretion disk images of the $\gamma$-metric,''
Eur. Phys. J. C \textbf{82}, no.8, 696 (2022).

\bibitem{Shaikh:2018lcc}
R.~Shaikh, P.~Kocherlakota, R.~Narayan and P.~S.~Joshi,
``Shadows of spherically symmetric blac kholes and naked singularities,''
Mon. Not. Roy. Astron. Soc. \textbf{482}, no.1, 52-64 (2019).

\bibitem{Shaikh:2019fpu}
R.~Shaikh,
``Black hole shadow in a general rotating spacetime obtained through Newman-Janis algorithm,''
Phys. Rev. D \textbf{100}, no.2, 024028 (2019).


\bibitem{Khodabakhshi:2020hny}
H.~Khodabakhshi, A.~Giaimo and R.~B.~Mann,
``Einstein Quartic Gravity: Shadows, Signals, and Stability,''
Phys. Rev. D \textbf{102}, no.4, 044038 (2020).

\bibitem{Wei:2019pjf}
S.~W.~Wei, Y.~C.~Zou, Y.~X.~Liu and R.~B.~Mann,
``Curvature radius and Kerr black hole shadow,''
JCAP \textbf{08}, 030 (2019).



\bibitem{Hennigar:2018hza}
R.~A.~Hennigar, M.~B.~J.~Poshteh and R.~B.~Mann,
``Shadows, Signals, and Stability in Einsteinian Cubic Gravity,''
Phys. Rev. D \textbf{97}, no.6, 064041 (2018).

\bibitem{Abdolrahimi:2015kma}
S.~Abdolrahimi, R.~B.~Mann and C.~Tzounis,
``Double Images from a Single black hole,''
Phys. Rev. D \textbf{92}, 124011 (2015).





\bibitem{Khodadi:2021gbc}
M.~Khodadi, G.~Lambiase and D.~F.~Mota,
``No-hair theorem in the wake of Event Horizon Telescope,''
JCAP \textbf{09}, 028 (2021).

\bibitem{Khodadi:2020jij}
M.~Khodadi, A.~Allahyari, S.~Vagnozzi and D.~F.~Mota,
``Black holes with scalar hair in light of the Event Horizon Telescope,''
JCAP \textbf{09}, 026 (2020).







\bibitem{Vagnozzi:2022moj}
S.~Vagnozzi, R.~Roy, Y.~D.~Tsai, L.~Visinelli, M.~Afrin, A.~Allahyari, P.~Bambhaniya, D.~Dey, S.~G.~Ghosh and P.~S.~Joshi, \textit{et al.}
``Horizon-scale tests of gravity theories and fundamental physics from the Event Horizon Telescope image of Sagittarius A$^*$,''
[arXiv:2205.07787 [gr-qc]].





\bibitem{Chen:2022nbb}
Y.~Chen, R.~Roy, S.~Vagnozzi and L.~Visinelli,
``Superradiant evolution of the shadow and photon ring of Sgr A$^\star$,''
Phys. Rev. D \textbf{106}, no.4, 043021 (2022).

\bibitem{Roy:2021uye}
R.~Roy, S.~Vagnozzi and L.~Visinelli,
``Superradiance evolution of black hole shadows revisited,''
Phys. Rev. D \textbf{105}, no.8, 083002 (2022).

\bibitem{Vagnozzi:2020quf}
S.~Vagnozzi, C.~Bambi and L.~Visinelli,
``Concerns regarding the use of black hole shadows as standard rulers,''
Class. Quant. Grav. \textbf{37}, no.8, 087001 (2020).





\bibitem{Vagnozzi:2019apd}
S.~Vagnozzi and L.~Visinelli,
``Hunting for extra dimensions in the shadow of M87*,''
Phys. Rev. D \textbf{100}, no.2, 024020 (2019).






\bibitem{Heydari-Fard:2021ljh}
M.~Heydari-Fard, M.~Heydari-Fard and H.~R.~Sepangi,
``Thin accretion disks around rotating black holes in 4$D$ Einstein\textendash{}Gauss\textendash{}Bonnet gravity,''
Eur. Phys. J. C \textbf{81}, no.5, 473 (2021).






\bibitem{Afrin:2021imp}
M.~Afrin, R.~Kumar and S.~G.~Ghosh,
``Parameter estimation of hairy Kerr black holes from its shadow and constraints from M87*,''
Mon. Not. Roy. Astron. Soc. \textbf{504}, 5927-5940 (2021).

\bibitem{Kumar:2020hgm}
R.~Kumar, S.~G.~Ghosh and A.~Wang,
``Gravitational deflection of light and shadow cast by rotating Kalb-Ramond black holes,''
Phys. Rev. D \textbf{101}, no.10, 104001 (2020).

\bibitem{Kumar:2018ple}
R.~Kumar and S.~G.~Ghosh,
``Black hole Parameter Estimation from Its Shadow,''
Astrophys. J. \textbf{892}, 78 (2020).


\bibitem{Kumar:2020owy}
R.~Kumar and S.~G.~Ghosh,
``Rotating black holes in $4D$ Einstein-Gauss-Bonnet gravity and its shadow,''
JCAP \textbf{07}, 053 (2020).

\bibitem{Atamurotov:2013sca}
F.~Atamurotov, A.~Abdujabbarov and B.~Ahmedov,
``Shadow of rotating non-Kerr black hole,''
Phys. Rev. D \textbf{88}, no.6, 064004 (2013).


\bibitem{Tang:2022hsu}
Z.~Y.~Tang, X.~M.~Kuang, B.~Wang and W.~L.~Qian,
``The length of a compact extra dimension from shadow,''
Sci. Bull. \textbf{67}, 2272-2275 (2022).

\bibitem{Kuang:2022ojj}
X.~M.~Kuang, Z.~Y.~Tang, B.~Wang and A.~Wang,
``Constraining a modified gravity theory in strong gravitational lensing and black hole shadow observations,''
Phys. Rev. D \textbf{106}, no.6, 064012 (2022).

\bibitem{Kuang:2022xjp}
X.~M.~Kuang and A.~\"Ovg\"un,
``Strong gravitational lensing and shadow constraint from M87* of slowly rotating Kerr-like black hole,''
Annals Phys. \textbf{447}, 169147 (2022).

\bibitem{Meng:2022kjs}
Y.~Meng, X.~M.~Kuang and Z.~Y.~Tang,
``Photon regions, shadow observables and constraints from M87* of a charged rotating black hole,''
Phys. Rev. D \textbf{106}, no.6, 064006 (2022).



\bibitem{Bambi:2015kza}
C.~Bambi,
``Testing black hole candidates with electromagnetic radiation,''
Rev. Mod. Phys. \textbf{89}, no.2, 025001 (2017).





\bibitem{Shakura:1972te}
N.~I.~Shakura and R.~A.~Sunyaev,
``Black holes in binary systems. Observational appearance,''
Astron. Astrophys. \textbf{24}, 337-355 (1973)

\bibitem{novikov}
 I. D. Novikov and K. S. Thorne, Astrophysics and black holes, in black holes, edited by C. De Witt
and B. De Witt (Gordon and Breach, New York, 1973).

\bibitem{Page:1974he}
D.~N.~Page and K.~S.~Thorne,
``Disk-Accretion onto a black hole. Time-Averaged Structure of Accretion Disk,''
Astrophys. J. \textbf{191}, 499-506 (1974).

\bibitem{Harko:2009xf}
T.~Harko, Z.~Kovacs and F.~S.~N.~Lobo,
``Thin accretion disks in stationary axisymmetric wormhole spacetimes,''
Phys. Rev. D \textbf{79}, 064001 (2009).

\bibitem{Torres:2002td}
D.~F.~Torres,
``Accretion disc onto a static nonbaryonic compact object,''
Nucl. Phys. B \textbf{626}, 377-394 (2002).

\bibitem{Perez:2012bx}
D.~Perez, G.~E.~Romero and S.~E.~P.~Bergliaffa,
``Accretion disks around black holes in modified strong gravity,''
Astron. Astrophys. \textbf{551}, A4 (2013).


\bibitem{Soroushfar:2020kgb}
S.~Soroushfar and S.~Upadhyay,
``Accretion disks around a static black hole in $f(R)$ gravity,''
Eur. Phys. J. Plus \textbf{135}, no.3, 338 (2020).

\bibitem{Pun:2008ae}
C.~S.~J.~Pun, Z.~Kovacs and T.~Harko,
``Thin accretion disks in f(R) modified gravity models,''
Phys. Rev. D \textbf{78}, 024043 (2008).

\bibitem{Staykov:2016dzn}
K.~V.~Staykov, D.~D.~Doneva and S.~S.~Yazadjiev,
``Accretion disks around neutron and strange stars in R+aR(2) gravity,''
JCAP \textbf{08}, 061 (2016).


\bibitem{demir-mdpi}
D.~Demir,
``Naturally-Coupled Dark Sectors,''
Galaxies \textbf{9}, no.2, 33 (2021).

\bibitem{demir-cankocak}
K.~Canko\c{c}ak, D.~Demir, C.~Karahan and S.~\c{S}en,
``Electroweak Stability and Discovery Luminosities for New Physics,''
Eur. Phys. J. C \textbf{80}, no.12, 1188 (2020).

\bibitem{Slany:2020jhs}
P.~Slan\'y and Z.~Stuchl\'\i{}k,
Eur. Phys. J. C \textbf{80} (2020) no.6, 587
doi:10.1140/epjc/s10052-020-8142-0

\bibitem{Gou:2011nq}
L.~Gou, J.~E.~McClintock, M.~J.~Reid, J.~A.~Orosz, J.~F.~Steiner, R.~Narayan, J.~Xiang, R.~A.~Remillard, K.~A.~Arnaud and S.~W.~Davis,
Astrophys. J. \textbf{742} (2011), 85
doi:10.1088/0004-637X/742/2/85
[arXiv:1106.3690 [astro-ph.HE]].

\bibitem{Dove:1997ei}
J.~B.~Dove, J.~Wilms, M.~Maisack and M.~C.~Begelman,
Astrophys. J. \textbf{487} (1997), 759
doi:10.1086/304647
[arXiv:astro-ph/9705130 [astro-ph]].


\end{thebibliography}
\end{document}